\documentclass[fdp,a4paper,fleqn%
]{w-art}
\usepackage{times,cite,w-thm}
\theoremstyle{plain}

\theoremstyle{definition}

\usepackage[]{graphicx}

\newcommand\rf[1]{(\ref{eq:#1})}
\newcommand\lab[1]{\label{eq:#1}}
\newcommand\nonu{\nonumber}
\newcommand\br{\begin{eqnarray}}
\newcommand\er{\end{eqnarray}}
\newcommand\be{\begin{equation}}
\newcommand\ee{\end{equation}}

\newcommand\llb{\left\lbrack}
\newcommand\rrb{\right\rbrack}

\newcommand\lcurl{\left\{}
\newcommand\rcurl{\right\}}
\renewcommand\({\left(}
\renewcommand\){\right)}
\newcommand\bv{\bigm\vert}               

\newcommand\bc{\begin{center}}
\newcommand\ec{\end{center}}

\newcommand\partder[2]{\frac{{\partial {#1}}}{{\partial {#2}}}}

\renewcommand\d{\delta}

\newcommand\vareps{\varepsilon}
\newcommand\g{\gamma}
\newcommand\G{\Gamma}

\newcommand\h{\frac{1}{2}}
\renewcommand\k{\kappa}
\renewcommand\l{\lambda}

\newcommand\m{\mu}
\newcommand\n{\nu}

\newcommand\vp{\varphi}
\renewcommand\P{\Phi}
\newcommand\pa{\partial}

\newcommand\pr{\prime}

\renewcommand\r{\rho}
\newcommand\s{\sigma}

\renewcommand\t{\tau}
\renewcommand\th{\theta}

\newcommand\wti{\widetilde}

\newcommand\cA{{\mathcal A}}

\newcommand\cF{{\mathcal F}}

\newcommand{\ct}[1]{\cite{#1}}
\newcommand{\bib}[1]{\bibitem{#1}}
\newcommand\PRL[3]{\textsl{Phys. Rev. Lett.} \textbf{#1}, #3 (#2)}

\newcommand\PRD[3]{\textsl{Phys. Rev.} \textbf{D#1}, #3 (#2)}

\newcommand\CQG[3]{\textsl{Class. Quantum Grav.} \textbf{#1}, #3 (#2)}

\newcommand\AoP[3]{\textsl{Ann. of Phys.} \textbf{#1}, #3 (#2)}

\newcommand\IJMPA[3]{\textsl{Int. J. Mod. Phys.} \textbf{A#1}, #3 (#2)}

\newcommand\rdot{\stackrel{.}{r}}

\begin{document}

\DOIsuffix{theDOIsuffix}
\Volume{}
\Issue{}
\Month{}
\Year{}
\pagespan{1}{}
\Receiveddate{}
\Reviseddate{}
\Accepteddate{}
\Dateposted{}



\title[Lightlike Branes and Wormholes]{Lightlike Branes as Natural Candidates for 
Wormhole Throats}


\author[F. Author]{E.I. Guendelman\inst{1,}\footnote{E-mail: 
~\textsf{guendel@bgumail.bgu.ac.il}.}}
\address[\inst{1}]{Department of Physics, Ben-Gurion University of the Negev,
P.O.Box 653\\ IL-84105 ~Beer-Sheva, Israel}
\author[S. Author]{A. Kaganovich\inst{1,}\footnote{E-mail: 
~\textsf{alexk@bgumail.bgu.ac.il}.}}
\author[T. Author]{E. Nissimov\inst{2,}
  \footnote{Corresponding author\quad E-mail:~\textsf{nissimov@inrne.bas.bg},
            Phone: +359\,2\,9795\,647,
            Fax: +359\,2\,975\,3619}}
\address[\inst{2}]{Institute for Nuclear Research and Nuclear Energy,
Bulgarian Academy of Sciences\\
Boul. Tsarigradsko Chausee 72, BG-1784 ~Sofia, Bulgaria}
\author[]{S. Pacheva\inst{2,}\footnote{E-mail: ~\textsf{svetlana@inrne.bas.bg}.}}

\begin{abstract}
We first briefly present a consistent world-volume Lagrangian description of 
lightlike $p$-branes (\textsl{LL-branes}) in two equivalent forms -- 
a Polyakov-type and a dual to it Nambu-Goto-type formulations. The most
important characteristic features of \textsl{LL-brane} dynamics are:
(i) the brane tension appears as a non-trivial additional dynamical degree of 
freedom; (ii) consistency of \textsl{LL-brane} dynamics in a spherically or axially
symmetric gravitational background of codimension one requires the presence
of an event horizon which is automatically occupied by the \textsl{LL-brane}
(``horizon straddling''). Next we consider a bulk Einstein-Maxwell system
interacting self-consistently with a codimension one \textsl{LL-brane}. We
find spherically symmetric traversable wormhole solutions of Misner-Wheeler type 
produced by the \textsl{LL-brane} sitting at the wormhole throat with wormhole
parameters being functions of the dynamical \textsl{LL-brane} tension.
\end{abstract}

\maketitle                   




\renewcommand{\leftmark}
{E.I. Guendelman et al.: Lightlike Branes and Wormholes}


\section{\label{sec:intro}Introduction}

The notion of wormhole space-time in the context of Schwarzschild geometry was 
first introduced by Einstein and Rosen \ct{Einstein-Rosen}
and understood in full details after the work of Kruskal and Szekeres 
\ct{Kruskal-Szekeres}. Einstein and Rosen \ct{Einstein-Rosen} also discussed 
the possibility of a singularity free wormhole solution obtained at the price 
of considering as source a Maxwell field with an energy momentum tensor being 
the opposite to the standard one. Misner and Wheeler \ct{Misner-Wheeler} also 
realized that wormholes connecting two asymptotically flat space times provide 
the possibility of ``charge without charge'', \textsl{i.e.}, electromagnetically
non-trivial solutions where the lines of force of the electric field flow from one 
universe to the other without a source and giving the impression of being 
positively charged in one universe and negatively charged in the other universe.

In order to construct wormholes of the traversable type, however, matter which 
violate the null energy condition must be introduced \ct{Visser-book}. An interesting 
consequence of this construction is that the resulting space time is non-singular.
Here we will discuss within this context the role of lightlike $p$-branes 
(\textsl{LL-branes} for short), whose explicit world-volume Lagrangian
formulation was given in our previous papers \ct{LL-branes-1}. As we will see in
what follows, \textsl{LL-branes} are particularly well suited to serve as 
gravitational sources of Misner-Wheeler type wormholes located at their throats, 
while at the same time they obey a well-defined dynamics derivable from first 
principles.

Let us note that \textsl{LL-branes} by themselves attract special interest in 
general relativity. This is due primarily because of their role in the
effective description of many cosmological and astrophysical effects such as
impulsive lightlike signals arising in cataclysmic astrophysical events
\ct{barrabes-hogan},  the ``membrane paradigm'' theory of black hole physics 
\ct{membrane-paradigm}, the thin-wall approach to domain walls coupled to gravity
\ct{Israel-66,Barrabes-Israel-Hooft}.
More recently \textsl{LL-branes} acquired significance also in the context of
modern non-perturbative string theory \ct{nonperturb-string}.

There are two basic properties of \textsl{LL-branes} which drastically
distinguish them from ordinary Nambu-Goto branes: (i) they describe
intrinsically lightlike modes, whereas Nambu-Goto branes describe massive
ones; (ii) the tension of the \textsl{LL-brane} arises as an additional
dynamical degree of freedom, whereas Nambu-Goto brane tension is a given
{\em ad hoc} constant.

\section{\label{sec:ll-brane}World-Volume Lagrangian Formulation of Lightlike Branes}

In refs.\ct{LL-branes-1,LL-branes-2} we have proposed a systematic Lagrangian 
formulation of a generalized Polyakov-type for \textsl{LL-branes} in terms of the 
world-volume action:
\be
S_{\mathrm{LL}} = \int d^{p+1}\s \,\P (\vp) \llb - \h \g^{ab} g_{ab} + 
L\!\( F^2\)\rrb \; .
\lab{LL-brane}
\ee
Here the following notations are used: $a,b=0,1,\ldots ,p$; 
$(\s^a)\equiv (\t,\s^i)$ with $i=1,\ldots ,p$; $\g_{ab}$ denotes the intrinsic 
Remain metric on the world-volume;
\be
g_{ab} \equiv \pa_a X^{\m} \pa_b X^{\n} G_{\m\n}(X)
\lab{ind-metric}
\ee
is the induced metric (the latter becomes {\em singular} on-shell -- lightlikeness, 
cf. second Eq.\rf{phi-gamma-eqs} below);
\be
\P (\vp) \equiv \frac{1}{(p+1)!} \vareps_{I_1\ldots I_{p+1}}
\vareps^{a_1\ldots a_{p+1}} \pa_{a_1} \vp^{I_1}\ldots \pa_{a_{p+1}} \vp^{I_{p+1}}
\lab{mod-measure-p}
\ee
is an alternative non-Riemannian reparametrization-covariant integration measure 
density replacing the standard $\sqrt{-\g} \equiv \sqrt{-\det \Vert \g_{ab}\Vert}$
and built from auxiliary world-volume scalars $\lcurl \vp^I \rcurl_{I=1}^{p+1}$;
\be
F_{a_1 \ldots a_{p}} = p \pa_{[a_1} A_{a_2\ldots a_{p}]} \quad ,\quad
F^{\ast a} = \frac{1}{p!} \frac{\vareps^{a a_1\ldots a_p}}{\sqrt{-\g}}
F_{a_1 \ldots a_{p}}
\lab{p-rank}
\ee
are the field-strength and its dual one of an auxiliary world-volume $(p-1)$-rank 
antisymmetric tensor gauge field $A_{a_1\ldots a_{p-1}}$ with Lagrangian $L(F^2)$ 
~($F^2 \equiv F_{a_1 \ldots a_{p}} F_{b_1 \ldots b_{p}} 
\g^{a_1 b_1} \ldots \g^{a_p b_p}$).

Equivalently one can rewrite \rf{LL-brane} as:
\br
S = \int d^{p+1}\!\!\s \,\chi \sqrt{-\g} 
\llb -\h \g^{ab} g_{ab} + L\!\( F^2\)\rrb \quad, \;\; 
\chi \equiv \frac{\P (\vp)}{\sqrt{-\g}}
\lab{LL-brane-chi}
\er
The composite field $\chi$ plays the role of a 
{\em dynamical (variable) brane tension}.

For the special choice $L\!\( F^2\)= \( F^2\)^{1/p}$ the above action  
becomes invariant under Weyl (conformal) symmetry:
$\g_{ab} \longrightarrow \g^{\pr}_{ab} = \rho\,\g_{ab}  \;\;,\;\;
\vp^{i} \longrightarrow \vp^{\pr\, i} = \vp^{\pr\, i} (\vp)$
with Jacobian $\det \Bigl\Vert \frac{\pa\vp^{\pr\, i}}{\pa\vp^j} \Bigr\Vert = \rho$. 

Consider now the equations of motion corresponding to \rf{LL-brane} 
w.r.t. $\vp^I$ and $\g^{ab}$:
\br
\h \g^{cd} g_{cd} - L(F^2) = M \quad , \quad
\h g_{ab} - F^2 L^{\pr}(F^2) \llb\g_{ab} 
- \frac{F^{*}_a F^{*}_b}{F^{*\, 2}}\rrb = 0  \; .
\lab{phi-gamma-eqs}
\er
Here $M$ is an integration constant and $F^{*\, a}$ is the dual field
strength \rf{p-rank}. Both Eqs.\rf{phi-gamma-eqs} imply the constraint
$L\!\( F^2\) - p F^2 L^\pr\!\( F^2\) + M = 0$, \textsl{i.e.}
\br
F^2 = F^2 (M) = \mathrm{const} ~\mathrm{on-shell} \; .
\lab{F2-const}
\er
The second Eq.\rf{phi-gamma-eqs} exhibits {\em on-shell singularity} 
of the induced metric \rf{ind-metric}: 
\br
g_{ab}F^{*\, b}=0 \; ,
\lab{on-shell-singular}
\er
\textsl{i.e.}, the tangent vector to the world-volume $F^{*\, a}\pa_a X^\m$
is {\em lightlike} w.r.t. metric of the embedding space-time.

Further, the equations of motion w.r.t. world-volume gauge field 
$A_{a_1\ldots a_{p-1}}$ (with $\chi$ as defined in \rf{LL-brane-chi} and
accounting for the constraint \rf{F2-const}) 
$~\pa_{[a}\( F^{\ast}_{b]}\, \chi\) = 0$
allow us to introduce the dual ``gauge'' potential $u$:
\be
F^{\ast}_{a} = \mathrm{const}\, \frac{1}{\chi} \pa_a u \; .
\lab{u-def}
\ee
We can rewrite second Eq.\rf{phi-gamma-eqs} (the lightlike constraint)
in terms of the dual potential $u$ as:
\be
\g_{ab} = \frac{1}{2a_0}\, g_{ab} - \frac{2}{\chi^2}\,\pa_a u \pa_b u \quad ,
\quad a_0 \equiv F^2 L^{\pr}\( F^2\)\bv_{F^2=F^2(M)} = \mathrm{const}
\lab{gamma-eqs-u}
\ee
($L^\pr(F^2)$ denotes derivative of $L(F^2)$ w.r.t. the argument $F^2$).
From \rf{u-def} and \rf{F2-const} we have the relation: 
\be
\chi^2 = -2 \g^{ab} \pa_a u \pa_b u \; ,
\lab{chi2-eq}
\ee
and the Bianchi identity $\nabla_a F^{\ast\, a}=0$ becomes:
\be
\pa_a \Bigl( \frac{1}{\chi}\sqrt{-\g} \g^{ab}\pa_b u\Bigr) = 0  \; .
\lab{Bianchi-id}
\ee

Finally, the $X^\m$ equations of motion produced by the \rf{LL-brane} read:
\br
\pa_a \(\chi \sqrt{-\g} \g^{ab} \pa_b X^\m\) + 
\chi \sqrt{-\g} \g^{ab} \pa_a X^\n \pa_b X^\l \G^\m_{\n\l}(X) = 0  \;
\lab{X-eqs}
\er
where $\G^\m_{\n\l}=\h G^{\m\k}\(\pa_\n G_{\k\l}+\pa_\l G_{\k\n}-\pa_\k G_{\n\l}\)$
is the Christoffel connection for the external metric.

It is now straightforward to prove that the system of equations 
\rf{chi2-eq}--\rf{X-eqs} for $\( X^\m,u,\chi\)$, which are equivalent to the 
equations of motion \rf{phi-gamma-eqs}--\rf{u-def},\rf{X-eqs} resulting from the 
original Polyakov-type \textsl{LL-brane} action \rf{LL-brane}, can be equivalently 
derived from the following {\em dual} Nambu-Goto-type world-volume action: 
\br
S_{\rm NG} = - \int d^{p+1}\s \, T 
\sqrt{- \det\Vert g_{ab} - \frac{1}{T^2}\pa_a u \pa_b u\Vert}  \; .
\lab{LL-action-NG}
\er
Here $g_{ab}$ is the induced metric \rf{ind-metric};
$T$ is {\em dynamical} tension simply related to the dynamical tension 
$\chi$ from the Polyakov-type formulation \rf{LL-brane-chi} as
$T^2= \frac{\chi^2}{4a_0}$ with $a_0$ -- same constant as in \rf{gamma-eqs-u}.

Henceforth we will consider the initial Polyakov-type form \rf{LL-brane} of the
\textsl{LL-brane} world-volume action. Invariance under world-volume 
reparametrizations allows to introduce the standard synchronous gauge-fixing 
conditions:
\be
\g^{0i} = 0 \;\; (i=1,\ldots,p) \; ,\; \g^{00} = -1
\lab{gauge-fix}
\ee
Also, in what follows we will use a natural ansatz for the ``electric'' part of the 
auxiliary world-volume gauge field-strength:
\be
F^{\ast i}= 0 \;\; (i=1,{\ldots},p) \quad ,\quad \mathrm{i.e.} \;\;
F_{0 i_1 \ldots i_{p-1}} = 0 \; .
\lab{F-ansatz}
\ee

\section{\label{sec:codim-one}Lightlike Branes in Gravitational Backgrounds:
Codimension One}

Let us consider codimension one \textsl{LL-brane} moving in a general spherically
symmetric background:
\be
ds^2 = - A(t,r)(dt)^2 + B (t,r) (dr)^2 + C(t,r) h_{ij}(\vec{\th}) d\th^i d\th^j \; ,
\lab{spherical-metric}
\ee
\textsl{i.e.}, $D=(p+1)+1$, with the simplest non-trivial ansatz for the 
\textsl{LL-brane} embedding coordinates $X^\m (\s)$: 
\be
t = \t \equiv \s^0 \;\; , \;\; r= r(\t) \;\; , \;\; \th^i = \s^i \;
(i=1,{\ldots},p) \; .
\lab{X-embed}
\ee
The \textsl{LL-brane} equations of motion \rf{gamma-eqs-u}--\rf{X-eqs}, taking into 
account \rf{gauge-fix}--\rf{F-ansatz}, acquire the form: 
\br
-A + B \rdot^2 = 0 \;\; ,\; \mathrm{i.e.}\;\; \rdot = \pm \sqrt{\frac{A}{B}}
\quad ,\quad
\pa_t C + \rdot \pa_r C = 0
\lab{r-const} \\
\pa_\t \chi + \chi \llb \pa_t \ln \sqrt{AB} 
\pm \frac{1}{\sqrt{AB}} \Bigl(\pa_r A + p\, a_0 \pa_r \ln C\Bigr)\rrb_{r=r(\t)} = 0
\; ,
\lab{X0-eq-1}
\er
where $a_0$ is the same constant appearing in \rf{gamma-eqs-u}.
In particular, we are interested in static spherically symmetric metrics in 
standard coordinates:
\be
ds^2 = - A(r)(dt)^2 + A^{-1}(r) (dr)^2 + r^2 h_{ij}(\vec{\th}) d\th^i d\th^j
\lab{standard-spherical}
\ee
for which Eqs.\rf{r-const} yield:
\be
\rdot = 0 \;\; ,\;\; \mathrm{i.e.}\;\; r(\t) = r_0 = \mathrm{const} \quad, \quad
A(r_0) = 0 \; .
\lab{horizon-standard}
\ee
Eq.\rf{horizon-standard} tells us that consistency of \textsl{LL-brane} dynamics in 
a spherically symmetric gravitational background of codimension one requires the 
latter to possess a horizon (at some $r = r_0$), which is automatically occupied 
by the \textsl{LL-brane} (``horizon straddling''). Further, Eq.\rf{X0-eq-1}
implies for the dynamical tension:
\be
\chi (\t) = \chi_0 
\exp\lcurl\mp \t \(\pa_r A\bv_{r=r_0} + \frac{2 p\, a_0}{r_0}\)\rcurl
\quad ,\quad \chi_0 = \mathrm{const} \; .
\lab{chi-eq-standard-sol}
\ee
Thus, we find a time-asymmetric solution for the dynamical
brane tension which (upon appropriate choice of the signs $(\mp)$ depending on the 
sign of the constant factor in the exponent on the r.h.s. of \rf{chi-eq-standard-sol})
{\em exponentially ``inflates'' or ``deflates''} for large times (for details 
we refer to the last two references in \ct{LL-branes-2}). This phenomenon is
an analog of the ``mass inflation'' effect around black hole horizons 
\ct{poisson-israel}.

Let us note, that similar results (``horizon straddling'' and ``inflation'' of the
brane tension) have been obtained also for \textsl{LL-brane} moving in
axially symmetric Kerr-Newman background \ct{LL-branes-3}.

\section{\label{sec:wormhole}Self-consistent Misner-Wheeler Traversable Wormhole 
Solutions}

Let us now consider a self-consistent bulk Einstein-Maxwell system free of 
electrically charged matter, coupled to a codimension one \textsl{LL-brane}:
\be
S = \int\!\! d^D x\,\sqrt{-G}\,\llb \frac{R(G)}{16\pi}
- \frac{1}{4} \cF_{\m\n}\cF^{\m\n}\rrb 
+ S_{\mathrm{LL}} \; .
\lab{E-M-LL}
\ee
Here $\cF_{\m\n} = \pa_\m \cA_\n - \pa_\n \cA_\m$ and $S_{\mathrm{LL}}$ is
the same \textsl{LL-brane} world-volume action as in \rf{LL-brane-chi}.
In other words, the \textsl{LL-brane} will serve as a gravitational source 
through its energy-momentum tensor (see Eq.\rf{T-brane} below).
The pertinent Einstein-Maxwell equations of motion read:
\be
R_{\m\n} - \h G_{\m\n} R =
8\pi \( T^{(EM)}_{\m\n} + T^{(brane)}_{\m\n}\) \quad, \quad
\pa_\n \(\sqrt{-G}G^{\m\k}G^{\n\l} \cF_{\k\l}\) = 0 \; ,
\lab{Einstein-Maxwell-eqs}
\ee
where $T^{(EM)}_{\m\n} = \cF_{\m\k}\cF_{\n\l} G^{\k\l} - G_{\m\n}\frac{1}{4}
\cF_{\r\k}\cF_{\s\l} G^{\r\s}G^{\k\l}$, 
and the \textsl{LL-brane} energy-momentum tensor is straightforwardly derived
from \rf{LL-brane-chi}:
\be
T^{(brane)}_{\m\n} = - G_{\m\k}G_{\n\l}
\int\!\! d^{p+1} \s\, \frac{\d^{(D)}\bigl(x-X(\s)\bigr)}{\sqrt{-G}}\,
\chi\,\sqrt{-\g} \g^{ab}\pa_a X^\k \pa_b X^\l  \; ,
\lab{T-brane}
\ee

Using \rf{T-brane} we will now construct a traversable {\em wormhole} solution to the 
Einstein equations \rf{Einstein-Maxwell-eqs} of Misner-Wheeler type 
\ct{Misner-Wheeler} following the standard procedure \ct{Visser-book}.
In other words, the \textsl{LL-brane} will serve as a gravitational source
of the wormhole by locating itself on its throat.

To this end let us take a spherically symmetric solution of \rf{Einstein-Maxwell-eqs}
of the form \rf{standard-spherical} in the absence of the \textsl{LL-brane}  
(\textsl{i.e.}, without $T^{(brane)}_{\m\n}$ on the r.h.s.), which possesses an 
event horizon at some $r=r_0$ (\textsl{i.e.}, $A(r_0)=0$). Consider now the
following modification of the metric \rf{standard-spherical}:
\be
ds^2 = - {\wti A}(\eta)(dt)^2 + {\wti A}^{-1}(\eta) (d\eta)^2 + 
(r_0 + |\eta|)^2 h_{ij}(\vec{\th}) d\th^i d\th^j \quad ,\quad
{\wti A}(\eta) \equiv A(r_0 + |\eta|) \;\;, 
\lab{spherical-WH}
\ee
where $-\infty < \eta < \infty$.
From now on the bulk space-time indices $\m,\n$ will refer to $(t,\eta,\th^i)$.
The new metric \rf{spherical-WH} represents two identical copies of the
exterior region ($r > r_0$) of the spherically symmetric space-time with
metric \rf{standard-spherical}, which are sewed together along the horizon
$r=r_0$. We will show that the new metric \rf{spherical-WH} is a solution of the full
Einstein equations \rf{Einstein-Maxwell-eqs}, {\em including} 
$T^{(brane)}_{\m\n}$ on the r.h.s.. Here the newly introduced coordinate $\eta$
will play the role of a radial-like coordinate normal w.r.t. the
\textsl{LL-brane} located on the horizon, which interpolates between two copies of
the exterior region of \rf{standard-spherical} (the two copies transform into
each other under the ``parity'' transformation $\eta \to - \eta$).

Inserting in \rf{T-brane} the expressions for $X^\m (\s)$ from
\rf{X-embed} and \rf{horizon-standard} and taking into account \rf{gamma-eqs-u},
\rf{gauge-fix}--\rf{F-ansatz} we get:
\be
T_{(brane)}^{\m\n} = S^{\m\n}\,\d (\eta)
\lab{T-S-0}
\ee
with surface energy-momentum tensor:
\be
S^{\m\n} \equiv - \frac{\chi}{(2a_0)^{p/2-1}}\,
\llb - \pa_\t X^\m \pa_\t X^\n + \g^{ij} \pa_i X^\m \pa_j X^\n 
\rrb_{t=\t,\,\eta=0,\,\th^i =\s^i} \quad , \;\; \pa_i \equiv \partder{}{\s^i} \; ,
\lab{T-S-brane}
\ee
where again $a_0$ is the integration constant parameter appearing in the 
\textsl{LL-brane} dynamics (cf. Eq.\rf{gamma-eqs-u}). Let us also note that 
unlike the case of test \textsl{LL-brane} moving in a spherically symmetric
background (Eqs.\rf{X0-eq-1} and \rf{chi-eq-standard-sol}), the dynamical brane 
tension $\chi$ in Eq.\rf{T-S-brane} is {\em constant}. This is due to the fact 
that in the present context we have a discontinuity in the Christoffel connection 
coefficients across the \textsl{LL-brane} sitting on the horizon ($\eta = 0$). 
The problem in treating the geodesic \textsl{LL-brane} equations of motion
\rf{X-eqs}, in particular -- Eq.\rf{X0-eq-1}, can be resolved following the approach 
in ref.\ct{Israel-66} (see also the regularization approach in ref.\ct{BGG},
Appendix A) by taking the mean value of the pertinent non-zero Christoffel 
coefficients across the discontinuity at $\eta = 0$. From the explicit form of
Eq.\rf{X0-eq-1} it is straightforward to conclude that the above mentioned
mean values around $\eta = 0$ vanish since now $\pa_r$ is replaced by $\pa/\pa \eta$,
whereas the metric coefficients depend explicitly on $|\eta|$. Therefore, in the 
present case Eq.\rf{X0-eq-1} is reduced to $\pa_\t \chi = 0$.

Let us now separate in \rf{Einstein-Maxwell-eqs}
explicitly the terms contributing to $\delta$-function singularities (these
are the terms containing second derivatives w.r.t. $\eta$, bearing in
mind that the metric coefficients in \rf{spherical-WH} depend on $|\eta|$):
\br
R_{\m\n} \equiv \pa_\eta \G^{\eta}_{\m\n} + \pa_\m \pa_\n \ln \sqrt{-G}
+ \mathrm{non-singular ~terms}
\nonu \\
= 8\pi \( S_{\m\n} - \h G_{\m\n} S^{\l}_{\l}\) \d (\eta) 
+ \mathrm{non-singular ~terms} \; .
\lab{E-M-eqs}
\er
The only non-zero contribution to the $\delta$-function singularities on
both sides of Eq.\rf{E-M-eqs} arises for $(\m\n)=(\eta \eta)$. In order to
avoid coordinate singularity on the horizon it is more convenient to
consider the mixed component version of the latter:
\be
R^{\eta}_{\eta} = 8\pi \( S^{\eta}_{\eta} - \h S^{\l}_{\l}\) \d (\eta) 
+ \mathrm{non-singular ~terms} \; .
\lab{E-M-eqs-eta}
\ee
Evaluating the l.h.s. of \rf{E-M-eqs-eta} through the formula: 
\be
R^r_r = - \h \frac{1}{r^{D-2}}\pa_r \( r^{D-2} \pa_r A\)
\lab{}
\ee
valid for any spherically symmetric metric of the form \rf{standard-spherical}
and recalling $r=r_0 + |\eta|$,
we obtain the following matching condition for the coefficients in front of
the $\delta$-functions on both sides of \rf{E-M-eqs-eta} (analog of Israel 
junction conditions \ct{Israel-66,Barrabes-Israel-Hooft}):
\be
\pa_\eta {\wti A}\bv_{\eta \to +0} - \pa_\eta {\wti A}\bv_{\eta \to -0} =
- \frac{16 \pi\,\chi}{(2a_0)^{p/2-1}} \quad, \;\; (\mathrm{recall} ~D=p+2) \; .
\lab{israel-junction}
\ee

Eq.\rf{israel-junction} yields a relation between the parameters of the
spherically symmetric outer regions of ``vacuum'' solution \rf{standard-spherical}
of Einstein Eqs.\rf{Einstein-Maxwell-eqs} and the dynamical tension of the 
\textsl{LL-brane} sitting at the (outer) horizon.

As an explicit example let us take \rf{standard-spherical} to be the
standard $D=4$ Reissner-Nordstr{\"o}m metric, \textsl{i.e.},
$A(r)= 1 - \frac{2M}{r} + \frac{e^2}{r^2}$. Then Eq.\rf{israel-junction}
yields the following relation between the Reissner-Nordstr{\"o}m
parameters and the dynamical \textsl{LL}-brane tension:
\be
4\pi\chi\,r_0^2 + r_0 - M = 0 \quad, \;\;\mathrm{where} \;\;
r_0 = M +\sqrt{M^2 -e^2} \; .
\lab{parameter-matching}
\ee
Eq.\rf{parameter-matching} indicates that the dynamical brane tension must
be {\em negative}. Eq.\rf{parameter-matching} reduces to a cubic equation
for the Reissner-Nordstr{\"o}m mass $M$ as function of $|\chi|$:
\be
\bigl( 16\pi\,|\chi|\, M - 1\bigr) \( M^2 - e^2\) + 16\pi^2 \chi^2 e^4 = 0 \; .
\lab{M-RN}
\ee
In the special case of Schwarzschild wormhole ($e^2 = 0$) the Schwazrschild mass
becomes:
\be
M = \frac{1}{16\pi\,|\chi|} \;\; .
\lab{M-Schw}
\ee
Let us observe that for large values of the \textsl{LL-brane} tension $|\chi|$, 
$M$ is very small. In particular, $M << M_{Pl}$ for $|\chi| > M_{Pl}^3$ 
($M_{Pl}$ being the Planck mass). On the other hand, for small values of the 
\textsl{LL-brane} tension $|\chi|$ Eq.\rf{parameter-matching} implies that the 
Reissner-Nordstr{\"o}m geometry of the wormhole must be near extremal 
($M^2 \simeq e^2$).

\section{\label{sec:conclude}Conclusions}

In the present note we have first stressed two fundamentally important properties 
of \textsl{LL-brane} dynamics: (a) the \textsl{LL-brane} tension appears as a
non-trivial additional dynamical degree of freedom instead of being given as
an \textsl{ad hoc} constant; (b) \textsl{LL-branes} of codimension one 
automatically locate themselves on horizons of space-times with black hole type 
geometries.

Next, we have proposed self-consistent spherically symmetric solutions to the
Einstein-Maxwell system which are traversable wormholes of Misner-Wheeler type and 
whose gravitational source are \textsl{LL-branes} located at their throats.
Specifically we have considered Reissner-Nordstr{\"o}m wormhole which is
built by sewing together two outer regions of Reissner-Nordstr{\"o}m
space-time along the external horizon automatically occupied by
the \textsl{LL-brane}, whose surface stress-energy tensor (derived from a
consistent world-volume action principle) implements the pertinent Israel
junction conditions. Thus, at the throat of the Reissner-Nordstr{\"o}m wormhole 
sits a neutral \textsl{LL-brane} with negative tension, the lines of force of 
the electric field go through the throat of the wormhole from one universe 
to the other, without any real electrical source anywhere, however in one universe
there is the appearance of a positive charge, while in the other there is the 
appearance of a negative charge.

Finally let us note that according to Eq.\rf{parameter-matching} (in particular 
Eq.\rf{M-Schw}) wormholes built from \textsl{LL-branes} with very high negative
tension have a small mass.

\vspace{.2in}
\textbf{Acknowledgments.}
E.N. and S.P. are supported by European RTN network
{\em ``Forces-Universe''} (contract No.\textsl{MRTN-CT-2004-005104}).
They also received partial support from Bulgarian NSF grants \textsl{F-1412/04}
and \textsl{DO 02-257}.
Finally, all of us acknowledge support of our collaboration through the exchange
agreement between the Ben-Gurion University of the Negev (Beer-Sheva, Israel) and
the Bulgarian Academy of Sciences.


\end{document}